\documentclass[reprint,twocolumn,amsmath,amssymb,aps,superscriptaddress]{revtex4-2}

\usepackage{graphicx}
\usepackage{dcolumn}
\usepackage{bm}
\usepackage{graphicx}
\usepackage{amssymb}
\usepackage{epstopdf, epsfig}
\usepackage{yfonts}
\usepackage[colorlinks=true, linkcolor=blue, citecolor=blue, urlcolor=blue]{hyperref}
\usepackage{fontenc}
\usepackage{upgreek}
\usepackage{mathrsfs}
\usepackage{booktabs}
\usepackage{pifont}
\usepackage{amsmath}
\usepackage{bm}
\usepackage{wasysym}
\usepackage{caption}
\usepackage{subcaption}
\usepackage{euscript}
\usepackage{natbib}
\usepackage{lipsum}
\usepackage{xcolor}

\usepackage{orcidlink}

\usepackage{subcaption}
\usepackage{ragged2e}
\DeclareCaptionJustification{justified}{\justifying}
\newcommand{\jfm}[1]{\textcolor{blue}{#1}}

\begin{document}

\title{Symmetric influence of forward and opposing tidal currents on rogue wave statistics}

\author{Saulo Mendes\,\orcidlink{0000-0003-2395-781X}}
\email{saulo.dasilvamendes@unige.ch}
\affiliation{Group of Applied Physics, University of Geneva, Rue de l'\'{E}cole de M\'{e}decine 20, 1205 Geneva, Switzerland}
\affiliation{Institute for Environmental Sciences, University of Geneva, Boulevard Carl-Vogt 66, 1205 Geneva, Switzerland}
\author{Ina Teutsch\,\orcidlink{0000-0001-8751-8096}}
\email{ina.teutsch@baw.de}
\affiliation{Helmholtz-Zentrum Hereon, Coastal Climate and Regional Sea Level Changes, Max-Planck-Stra{\ss}e 1, 21502 Geesthacht, Germany}
\affiliation{Federal Waterways Engineering and Research Institute (BAW), Hamburg, 22559, Germany}
\author{Jérôme Kasparian\,\orcidlink{0000-0003-2398-3882}}
\email{jerome.kasparian@unige.ch}
\affiliation{Group of Applied Physics, University of Geneva, Rue de l'\'{E}cole de M\'{e}decine 20, 1205 Geneva, Switzerland}
\affiliation{Institute for Environmental Sciences, University of Geneva, Boulevard Carl-Vogt 66, 1205 Geneva, Switzerland}

\begin{abstract}
Rogue waves are associated with various ocean processes, both at the coast and in the open ocean. In either zone,
inhomogeneities in the wave field caused by shoaling, crossing seas or current interactions disturb wave statistics, increasing rogue wave probability and magnitude. Such amplification of the frequency of rogue waves and their intensity, i.e. the maximum normalized height, have been attested in numerical simulations and laboratory studies, in particular for wave-current interactions. In this study, we investigate the effect of the current intensity and direction on rogue wave probability, by analyzing long-term observations from the southern North Sea. 
We observe that the amplification is similar for opposing and following currents. Despite the sea states being dominantly broad-banded and featuring a large directional spread, anomalous statistics are of the same order of magnitude as those observed in unidirectional laboratory experiments for stationary currents.
\end{abstract}


\maketitle

\section{INTRODUCTION}

Waves exceptionally higher than the mean of the highest surrounding ones, so-called rogue waves, have been extensively observed over two decades by buoys \citep{Liu2004b,Doong2010,Srokosz2018,Hafner2021},
satellite imagery \citep{Rosenthal2007,Rosenthal2008}, oil platform sensors \citep{Haver2000b,Stansell2004,Mendes2021a} or a combination thereof \citep{Christou2014,Teutsch2020,Ewans2020,Teutsch2023}. They are generally defined as waves exceeding twice the significant wave height $H_s$, which is the mean height of the largest third of the waves in a given sea state. Accidents near the coast of South Africa half a century ago due to the interaction of wave trains traveling in opposition to strong surface currents provided substantial evidence of their existence  \citep{Mallory1974,Smith1976,Lavrenov1998}. However, they were disregarded, since their existence challenged the linear theory of irregular ocean waves. 

Waves in inhomogeneous media have been investigated from a deterministic point of view since the 1940's \citep{Unna1942,Johnson1947}. Advanced mathematical techniques \citep{Ursell1960,Whitham1962,Whitham1965,Bretherton1968} allowed to assess wave-current interactions through ray theory \citep{Arthur1950,Whitham1960,Kenyon1971}, linear wave theory \citep{Taylor1955,Peregrine1976}, radiation stress \citep{Higgins1960,Higgins1961}, and spectral \citep{Huang1972} as well as perturbative methods \citep{McKee1974}. The works of \citet{Higgins1952,Higgins1963} on the statistics of interaction-free water waves were conducted in parallel with his work on currents \citep{Higgins1960},
yet no work extended these statistical analyses to wave-current systems. Although the possibility of the existence of rogue waves was originally raised from wave-current observation \citep{Mallory1974}, the different focus placed by communities working on wave statistics on one side and on theoretical fluid dynamics of wave-current systems on the other side persisted for decades.
In fact, the study of how currents affect wave statistics and extreme events came under consideration only recently \citep{White1998,Heller2008,Janssen2009}.
However, the ray theory of wave-current statistics \citep{Heller2008,Heller2011} has not been tested against observations or laboratory experiments. Due to their definition relative to the significant wave height rather than an absolute height, the occurrence of rogue waves is independent of the absolute value of any sea state variable, such as significant wave height or mean wavelength \citep{Stansell2004,Christou2014,Mendes2021a}. Therefore, effects of the currents on either the wave height or wavelength cannot be translated into modulations of the rogue wave probability.

Focusing on the consequences of the interaction of waves with currents on wave height statistics, a proper theoretical study of both wave groups and random wave fields using the Nonlinear Schr\"{o}dinger equation (NLSE) showed good agreement with numerical simulations on the maximum amplitude driven by the interaction of the wave train with opposing currents \citep{Onorato2011}. Moreover, \citet{Toffoli2013} computed the maximum wave amplitude from the NLSE and found good agreement with unidirectional experiments with opposing currents carried out in two different wave flumes. Further experiments were performed to investigate broad-banded waves with finite directional spreading, demonstrating a decrease in amplification with directionality \citep{Toffoli2015}. Experiments by \citet{Ducrozet2021} have demonstrated that the amplification of the rogue wave probability due to an opposing current is related to the magnitude of the ratio between current speed and wave group speed $U/c_g$. 

\begin{figure*}
\centering
    \includegraphics[scale=0.45]{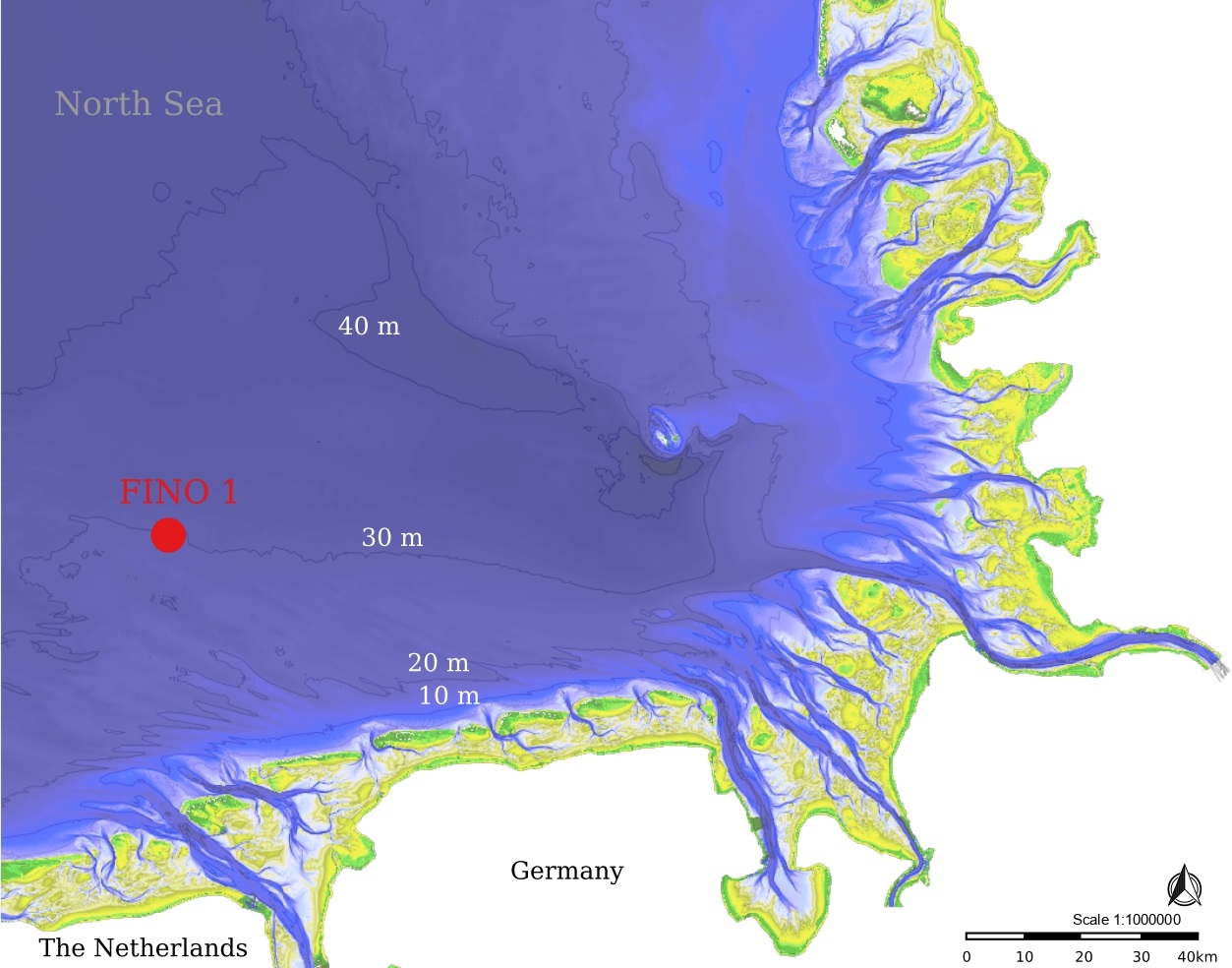}
    \caption{Location of the research platform FINO1 in the southern North Sea, close to the Dutch and German Frisian islands. Data from~\cite{MDIDE}.}  
    \label{fig:FINO}
\end{figure*}
When dealing with following currents, studies on either wave modulation or wave statistics are scarce. \citet{Trulsen2009} were the first to provide evidence that not only opposing, but also following currents may increase rogue wave occurrence. Recently, simulations performed by \citet{Li2023} and flume experiments by \citet{Benoit2023} have upheld the amplification influence of following currents, although the latter included depth effects of waves traveling over a bar. \citet{Benetazzo2024} analyzed the effect of non-stationary (and/or) non-homogeneous following and opposing currents (or tides) on the maximum height of irregular wave fields, but they did not investigate the probability distributions of wave heights.
Generalizing these studies to the wave height distribution, our present work aims to investigate how the interaction of waves with both opposing and following currents amplifies rogue wave occurrence under real ocean conditions, i.e. waves of broad spectrum and directional spreading subject to interaction with tidal (non-stationary) currents in the southern North Sea.

\section{Data and methods}\label{sec:methods}

As described in detail in \citet{TeutschSMJK2023}, wave and current data were recorded in the southern North Sea (FINO~1 research platform, 54.015$^{\circ}$N 6.588$^{\circ}$E, 30~m depth), see figure~\ref{fig:FINO}, between July 2019 and December 2022. Tidal currents are typically east- and westward, while the waves (of peak frequency 0.11--0.14~Hz) propagated mostly south-southeast- or eastward. Surface elevation was measured at a rate of 1.28~Hz and with a resolution of 0.01~m, in 30~minute long samples. An analysis of the sea conditions showed that stronger tidal currents symmetrically decrease the wavelength and bandwidth and increase the directional spread \citep{TeutschSMJK2023}. Furthermore, moderate currents boost the significant wave height while opposing currents push up the wave steepness.

\begin{table*}
\centering
\begin{tabular}{cccccccccccc}    
\toprule
\emph{Pairs $U/c_{g}$}  &   \emph{$(-0.19, 0)$} & \emph{$(-0.13, 0)$} & \emph{$(-0.06, 0)$} & \emph{$(+0.06, 0)$} & \emph{$(+0.13, 0)$} & \emph{$(+0.21, 0)$} 
\\
\midrule
$p$-value  & 0.293  & \quad 0.027 & \quad 0.042 & \quad 0.023 & \quad 0.015  &  0.053
\\
\\
\emph{Pairs $U/c_{g}$}  &   \emph{$(-0.06, +0.06)$} & \emph{$(-0.13, +0.13)$} & \emph{$(-0.19, +0.21)$} 
\\
\midrule
$p$-value  & 0.048   &  0.091  &  0.002 
\\
\bottomrule
\end{tabular}
\caption{Hypergeometric $p$-value calculated from the \citeauthor{Fisher1922}'s exact test for all possible pairs of data points of \jfm{figure} \ref{fig:URogueFisher}.} 
\label{tab:2x4}
\end{table*}
The vertical current profile was measured with a Nortek acoustic Doppler current profiler (ADCP), quality-controlled, and made available to us by the German Federal Maritime and Hydrographic Agency (BSH). Unless otherwise specified, we considered in our analyses the current at 5.5~m water depth. As detailed in Section \ref{section:SensitivityDepth}, we checked that, due to the smooth measured current profiles \citep{Teutsch2023} and the recorded wavelengths (peak wavelength $>$100~m) largely exceeding the water depth of 30~m, the choice of water depth does not influence the results. 

We defined following and opposing currents as currents within 
$\pm 10^{\circ}$ of the 0 and $180^{\circ}$ angle between current and peak wave frequency, respectively. This way, the cosine of the angle range lies beyond $\pm 0.98$ and the sine keeps below $\pm 0.17$: the transverse flow is negligible and the longitudinal flow is not affected by the slight deviations from the wave axis. The full data set gathers 4686 30-minute long samples, each containing 366~waves on average: 2156 of these with opposing current, 2321 with forward current, and 209 in rest conditions, defined as $|U/c_{g}| \lesssim 0.02$, in agreement with \citet{TeutschSMJK2023}. We checked that our results are robust against the choice of the current threshold defining rest conditions, as described in Section~\ref{section:SensitivityDepth}.

Large ship wakes may contaminate the data by producing a few large waves that could affect the tail of the wave height distribution, in which we were specifically interested. Existing methods for spotting ship wakes in time series~\citep{Didenkulova2015}, based on outlier detection, were not applicable due to our focus on the tail of the distribution. Therefore, we identified data that may be distorted by passing ships, relying on Automatic Identification System (AIS) recordings obtained from Marine Traffic, that provided ship length, heading, and velocity. The half angle of the wake, known as the Kelvin wedge \citep{Soomere2007} amounts to $\theta\approx 19^{\circ}$~\citep{Havelock1908}. Its length $L_x$ is proportional to the Froude number $Fr$ \citep{Thomson1887,Havelock1908} and can be estimated as
$L_{x} \approx (\ell/5)  (Fr)^{3/2} \approx 200 \ell (\mathcal{U}/h)^{3/2}$, where $\mathcal{U}$ is the ship velocity, $\ell$ the ship length and $h$ the water depth \citep{Voropayev2012,Rovelli2016}.
Large ships ($\ell \ge$ 300~m) traveling at $\mathcal{U}$ = 15 knots ($\sim$ 8 m/s) yield a wake with a length  $L_{x} \sim 8$~km  and a half width up to $L_{y} = 8~\textrm{km} \cdot \tan{(19^{\circ})} \approx 3$~km. We therefore excluded all 30-minute samples during which a ship passed by within 3~km from the platform. This corresponds to 19 samples, i.e. 9.5~hours or 0.03\% of the entire data set.

\section{Results} \label{sec:results}

\subsection{Rogue wave amplification is symmetric with regard to current direction}

\begin{figure}
    \centering
    \includegraphics[scale=0.45]{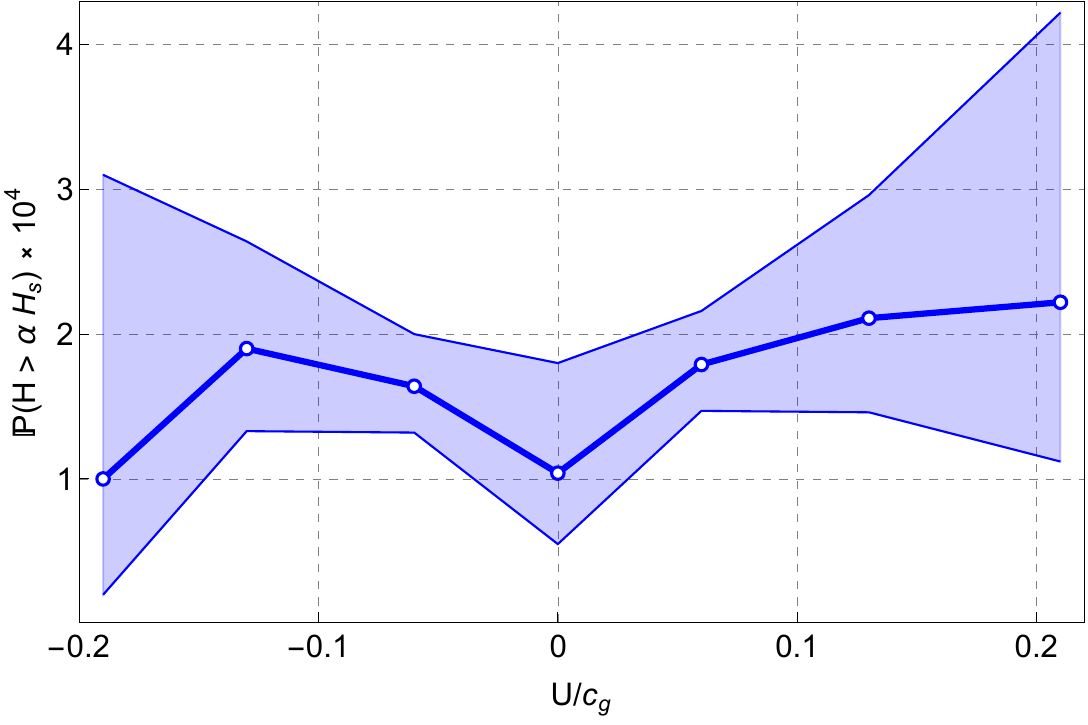}
    \caption{Exceedance probability of rogue waves ($\alpha = 2.0$) in the southern North Sea with low-resolution binning of the normalized tidal current $U/c_{g}$. The error band is computed from the 95\% \citeauthor{Jeffreys1961} confidence interval.}
    \label{fig:URogueFisher}
\end{figure}
To characterize the tail of the wave height distribution, we focus on the exceedance probability $\mathbb{P}(H> \alpha H_s)$, $\alpha$ being the normalized wave height. For rogue waves, $\mathbb{P}(\alpha = 2)$ is enhanced by moderate currents of either direction, whether following or opposing (\jfm{figure}~\ref{fig:URogueFisher}). At larger current velocity, this amplification saturates or decays. The limited number of rogue waves (278 events in the entire sample after quality control) however limits the resolution of the binning, preventing us from accurately determining the current speed associated with maximum rogue wave amplification. This limited number also explains the width of the shaded 95\% confidence ranges in \jfm{figure}~\ref{fig:URogueFisher}, which were calculated by assuming independent draws for each wave and consequently using a binomial analysis \citep[e.g]{Jeffreys1961}. The statistical significance of the results was further assessed by performing pairwise comparisons between bins with the \citet{Fisher1922} exact test, thereby computing their bilateral hypergeometric $p$-value (\jfm{table}~\ref{tab:2x4}). The two bins with smaller current on either direction ($U/c_g = \pm 0.06, \pm 0.13$) display significantly higher exceedance probabilities than the rest condition ($p < 0.05$). In contrast, the corresponding pairs of bins with the same velocity magnitude but opposing directions ($U/c_g = -0.06$ and 0.06, respectively, as well as $U/c_g = -0.13$ and 0.13) are respectively at the very edge of significance and non-significant. This analysis confirms the amplification of the rogue wave probability by a current, and the symmetry of this amplification with regard to the current direction, at least for moderate currents velocities. Conversely, the limited number of events limits the significance of the results for the outermost bin ($U/c_g = -0.19$ and $U/c_g = 0.21$). 
\begin{figure}
\vspace{0.0cm}
\centering
    \includegraphics[scale=0.45]{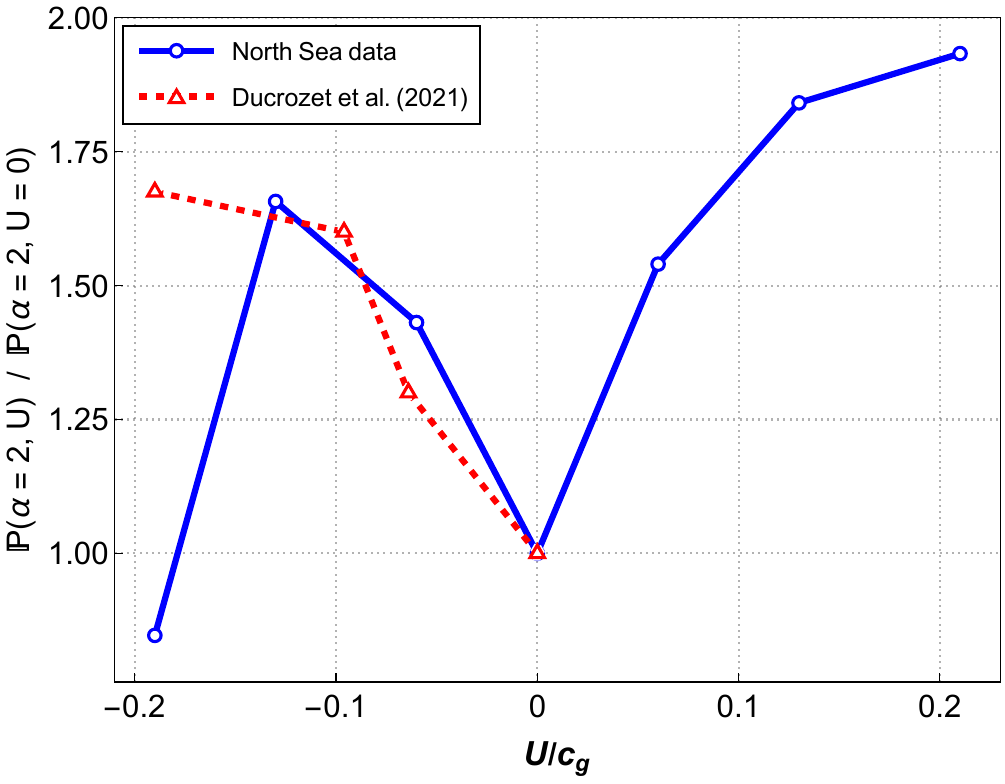}
\caption{Amplification of the rogue wave exceedance probability by opposing and following currents for the full southern North Sea dataset in comparison with laboratory measurements of \citet{Ducrozet2021}, eq.~(\ref{eq:2+1}).}
\label{fig:Phom}
\end{figure}
\begin{figure*}
\hspace{0.7cm}
\minipage{0.4\textwidth}
    \includegraphics[scale=0.5]{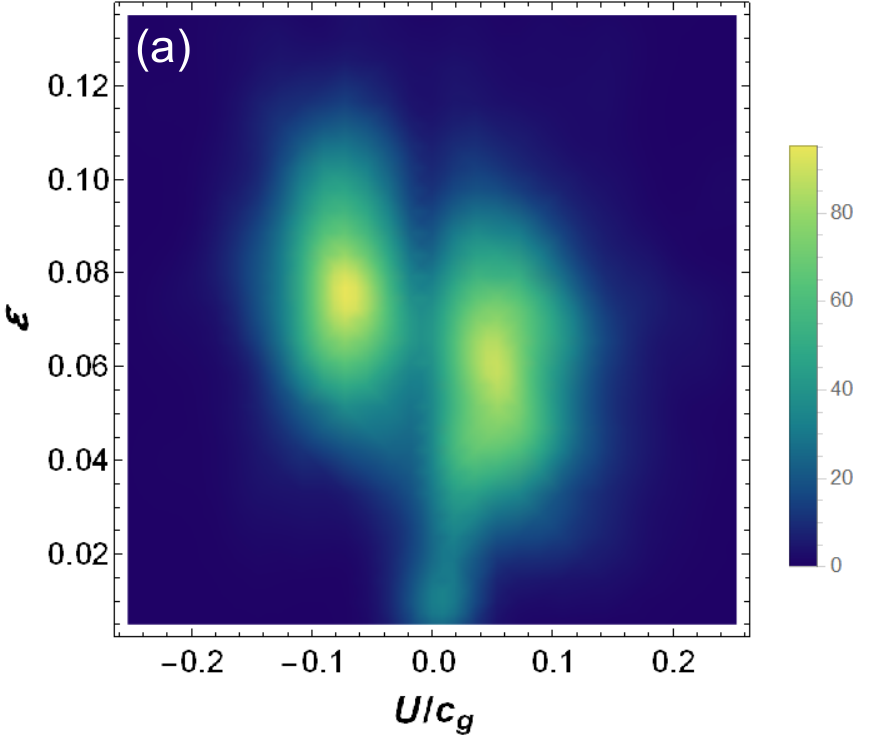}
\endminipage
\hfill
\minipage{0.54\textwidth}
    \includegraphics[scale=0.49]{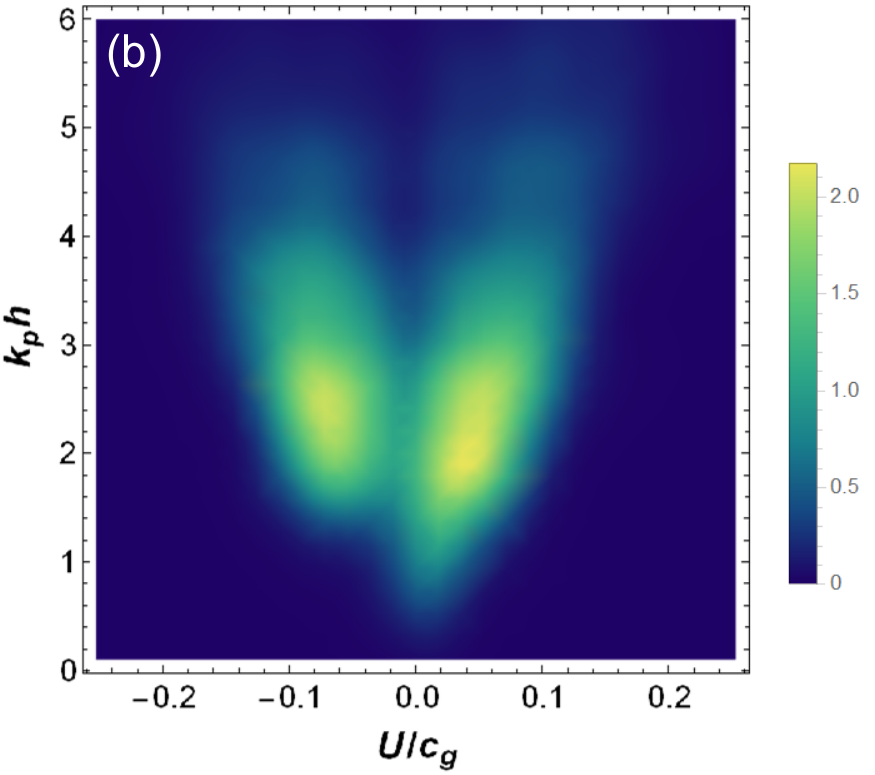}
\endminipage

\vspace{0.3cm}
\hspace{0.7cm}
\minipage{0.4\textwidth}
    \includegraphics[scale=0.5]{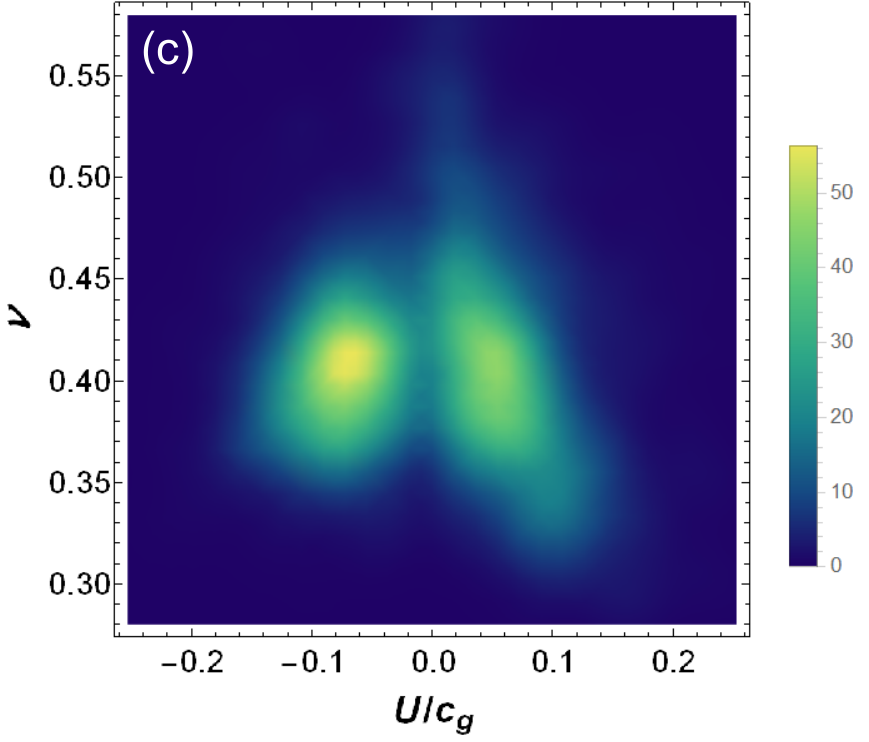}
\endminipage
\hfill
\minipage{0.54\textwidth}
    \includegraphics[scale=0.49]{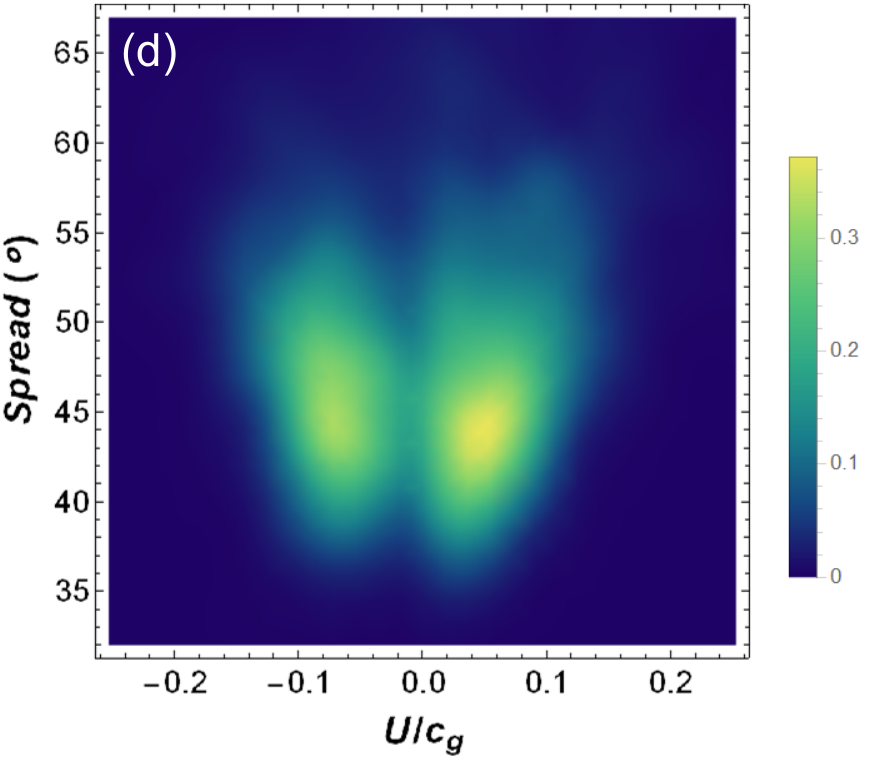}
\endminipage
\caption{Joint probability densities of several sea parameters and normalized currents speed $U/c_{g}$: (a) mean wave steepness $\varepsilon = (\sqrt{2}/\pi) k_p H_s$, (b) relative water depth $k_p h$, (c) bandwidth $\nu$ \citep{Higgins1975}, (d) directional spread $\sigma_{\theta} = \sqrt{2(1+s)}$ for a directional spectral function $D(\theta) \sim \cos^{2s}{(\theta/2)}$.}
\label{fig:spectraldist2}
\end{figure*}

\jfm{Figure}~\ref{fig:Phom} compares the rogue wave probability amplification observed in our data with \citet{Ducrozet2021}'s experimental results.
The experiments are one-dimensional (1+1, unidimensional in space and time). In contrast, in our data waves show a significant directional spread, introducing a second spatial dimension (2+1). Therefore, the exceedance probabilities in rest conditions are different, so that we compared the amplification ratios with regard to the latter:
\begin{equation}
    \frac{\mathbb{P}_{U/c_g}^{(2+1)}(\alpha )}{\mathbb{P}_{U/c_g=0}^{(2+1)}(\alpha )} \quad \textrm{and} \quad  \frac{\mathbb{P}_{U/c_g}^{(1+1)}(\alpha )}{\mathbb{P}_{U/c_g=0}^{(1+1)}(\alpha )} \quad .
\label{eq:2+1}
\end{equation}
The match between field observations and flume experiments is remarkable, especially when keeping in mind the wide range of conditions and the directional spread present in our dataset.

\subsection{Influence of the threshold $\alpha$ for the exceedance probabilities}

\begin{figure}
\centering
    \includegraphics[scale=0.45]{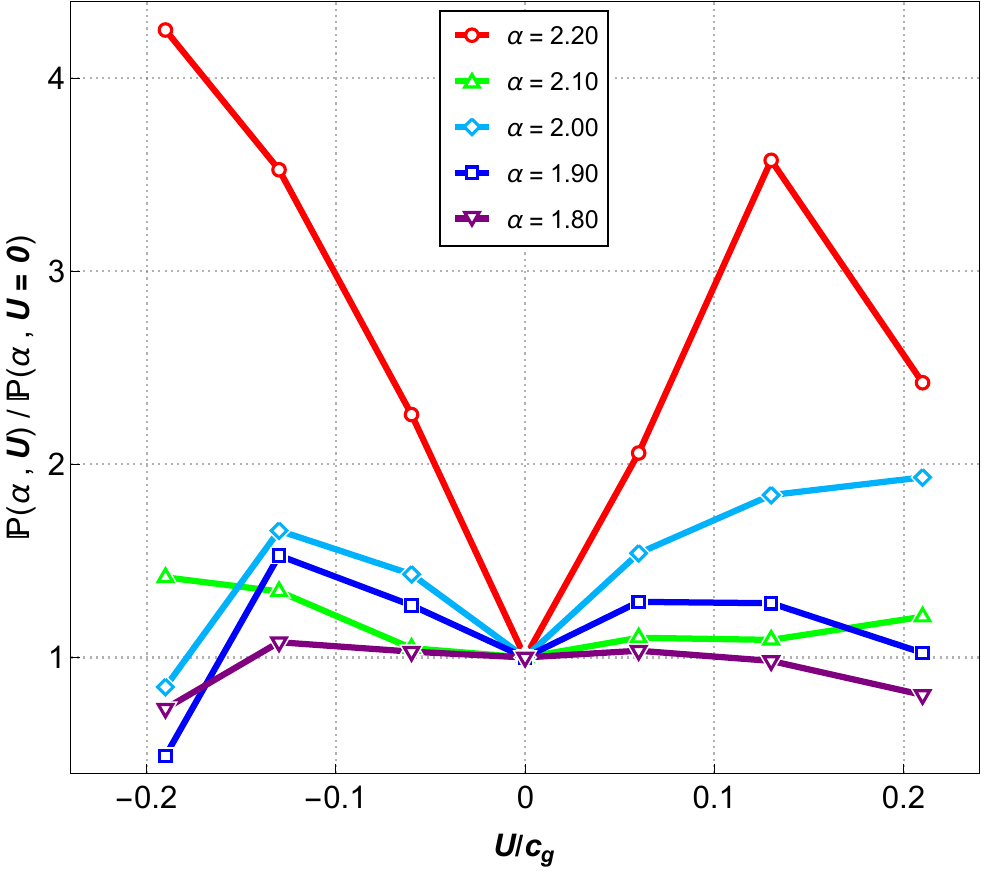}
\caption{Amplification of the exceedance probability of extreme waves as a function of the current velocity, for several values of the threshold $\alpha$.} 
\label{fig:Ualpha2b}
\end{figure}
Hitherto, we have focused on the exceedance probability for a value of $\alpha = 2$, corresponding to the common definition of rogue waves. As displayed in \jfm{figure}~\ref{fig:Ualpha2b}, the choice of $\alpha$ does not qualitatively impact our main finding, as long as $\alpha$ is sufficiently high. For all $\alpha \ge 1.9$, the exceedance probability is amplified by both following and opposing currents, up to a normalized current speed $|U/c_g|\approx 0.13$. The amplification tends to increase for larger values of $\alpha$, where a smaller fraction of the wave height distribution is included, enhancing the weight of the tail of the probability distribution function.

\subsection{Influence of data inhomogeneity}

Unlike flume experiments, our observations cover a wide range of sea states and cannot be considered homogeneous (\jfm{figure}~\ref{fig:spectraldist2}). 
Such sample variability is known to affect wave statistics \citep{Bitner2018}. 
To assess the effect of this inhomogeneity, 
we selected homogeneous subsets of our data with ranges in bandwidth, directional spread, and wave steepness close to the peak of the probability distribution of each parameter. We selected a steepness range of $\varepsilon \in [0.06,0.08]$, a bandwidth $\nu \in [0.35,0.45]$, as well as a directional spread $\in [45^\circ, 55^\circ$]. These conditions respectively select 27\%, 57\%, and 47\% of the sample (10\% when taken together) in the rest conditions, and 35\%, 90\%, and 64\% (24\% alltogether) for the bin of highest amplification (labeled above as $U/c_g = -0.13$, and corresponding to the range $U/c_g \in [-0.16,-0.10]$). Furthermore, we restricted ourselves to deep water, in order to minimize the effect of the reduced depth $k_p h$ on wave statistics. 
\begin{figure}
\centering
    \includegraphics[scale=0.4]{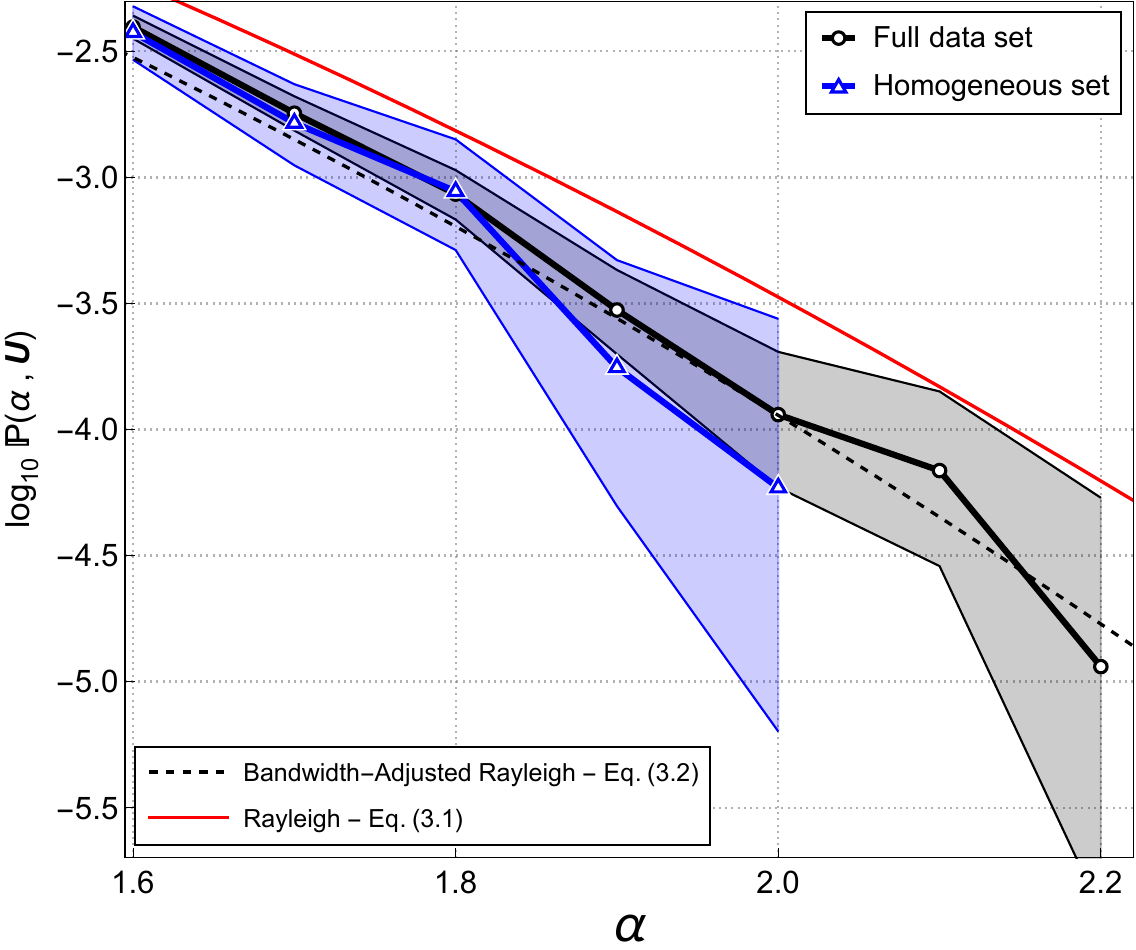}
\caption{Exceedance probability of heights exceeding the threshold $\alpha H_{s}$ in rest conditions observed in the North Sea data. Error bands were computed with 95\% \citeauthor{Jeffreys1961}'s confidence intervals.}
\label{fig:Ualphaexc}
\end{figure}

In the rest conditions, the exceedance probabilities for the full and the homogeneous datasets are similar up to $\alpha = 1.8$ (\jfm{figure}~\ref{fig:Ualphaexc}). Both lie below the Rayleigh exceedance probability~\citep{Higgins1952},
\begin{equation}
\mathbb{P}(\alpha) \equiv \mathbb{P}(H > \alpha H_{s}) = e^{-2\alpha^{2}} \quad ,
\label{eq:P1}
\end{equation}
and tend to follow the bandwidth-adjusted Rayleigh distribution~\citep{Higgins1980,Mendes2020}:
\begin{equation}
\mathbb{P}_{\nu}(\alpha)  = e^{-2\nu_{\star}\alpha^{2}} \quad , \quad \nu_{\star} = \left[ 1 - \left( \frac{\pi^2}{8} - \frac{1}{2} \right) \nu^{2} \right]^{-1} \quad .
\label{eq:P2}
\end{equation}
For $\alpha \ge 1.9$, the full dataset further follows the bandwidth-corrected Rayleigh distribution, while the homogeneous dataset tends to deviate towards lower exceedance probabilities. This exceedance probability lower than the broadband-corrected Rayleigh distribution is likely due to the suppression of the samples with the narrowest directional spreads, see for instance eq.~(22,24) of \citet{Mori2011} or the parameterization of \citet{Karmpadakis2022}. This decrease in probability also reduces the maximum normalized wave height \citep{Gumbel1958,Benetazzo2015,Mendes2021a} by 10\%.
\begin{figure}
\centering
    \includegraphics[scale=0.46]{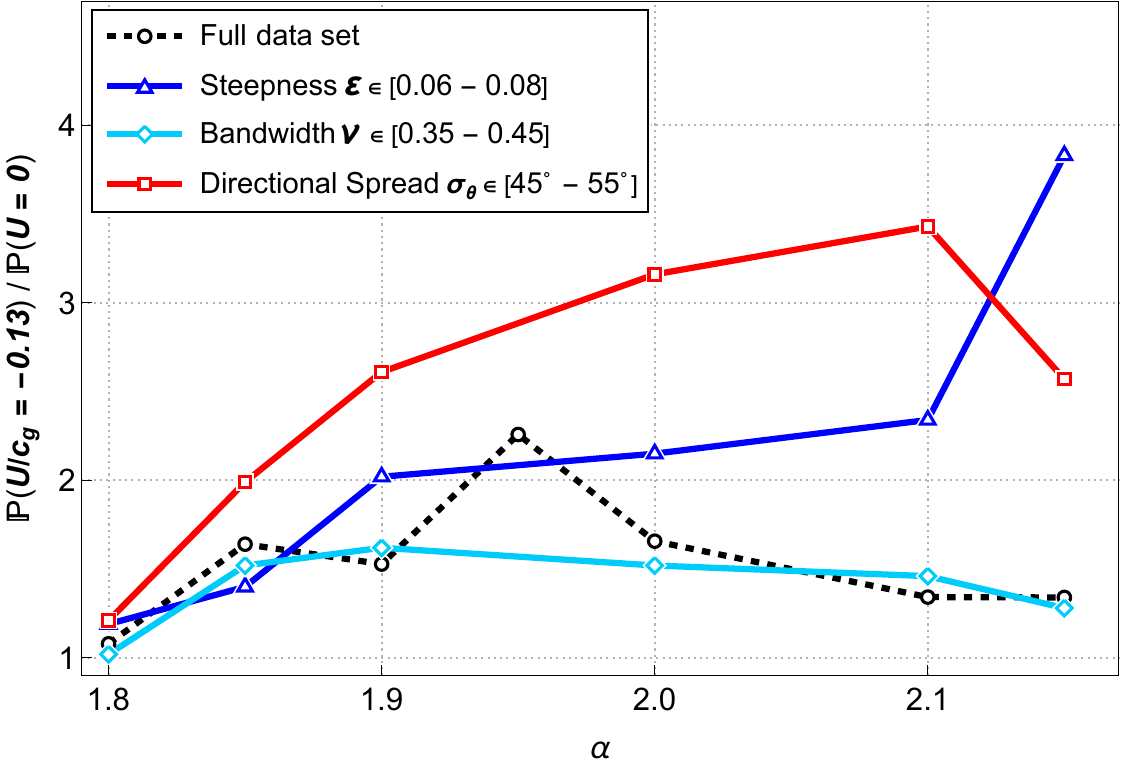}
\caption{Peak of amplification as a function of the normalized height threshold $\alpha$.} 
\label{fig:Ualpha2d}
\end{figure} 

\jfm{Figure}~\ref{fig:Ualpha2d} displays the peak of the rogue wave probability amplification (corresponding to $U/c_g = -0.13$) as a function of $\alpha$. It compares the full dataset with its counterparts obtained when individually selecting each of the the above-described homogeneous ranges of steepness, bandwidth, and directional spread. Narrowing the bandwidth range has no effect on the amplification. Conversely, selecting the data subset with a homogeneous steepness increases the amplification beyond $\alpha \gtrsim 2.0$. The effect is even stronger for the directional spread, as it starts already for $\alpha \gtrsim 1.85$, and reaches a factor of 2 or more as compared to the full dataset. Combining the selections of steepness, bandwidth, and directional spread further increases the amplification, although the results are statistically weaker, as the amount of samples drastically decreases. Therefore, they are not presented here. The up to twice higher amplification in the homogeneous datasets corresponds to an actual increase of the exceedance probabilities: in the homogeneous dataset, the exceedance probability amplification is at most 40\% lower than in the rest conditions (See \jfm{figure}~\ref{fig:Ualphaexc}).

\subsection{Influence of nonlinearity}

\begin{figure}
\centering
    \includegraphics[scale=0.46]{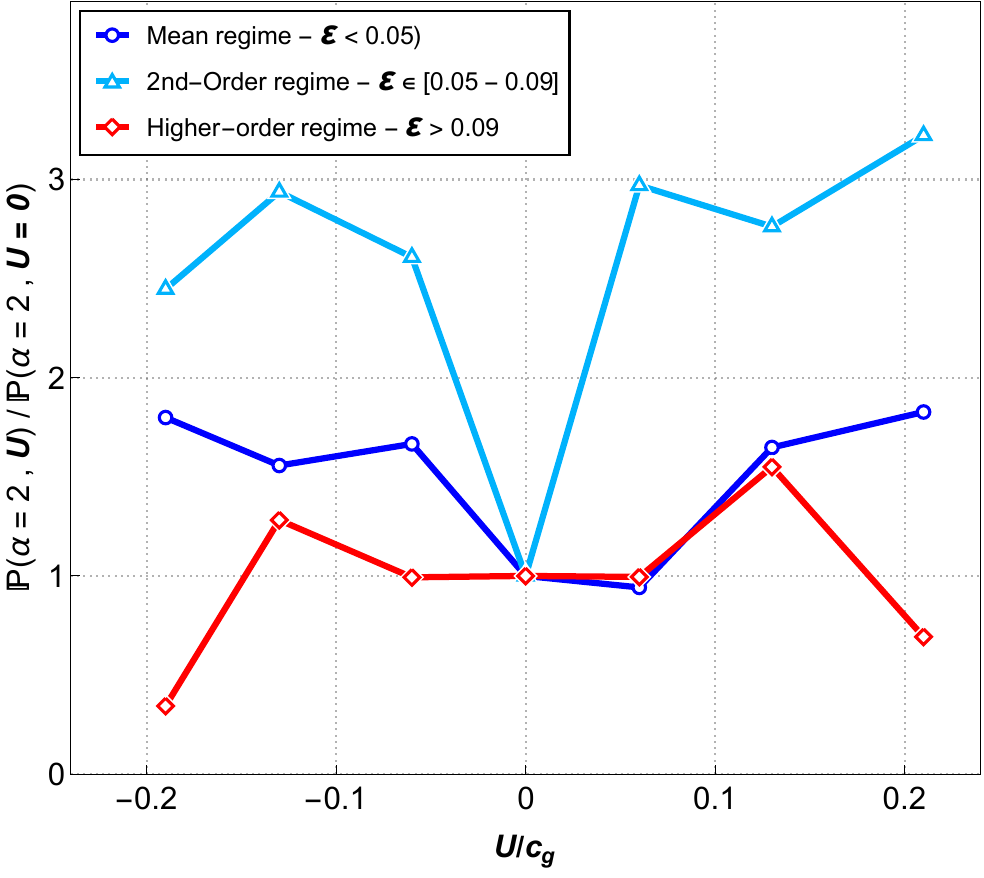}
\caption{Effect of steepness on the rogue wave amplification by tidal currents for different linearity ranges.} 
\label{fig:Ualpha3}
\end{figure}
\begin{figure}
\centering
    \includegraphics[scale=0.45]{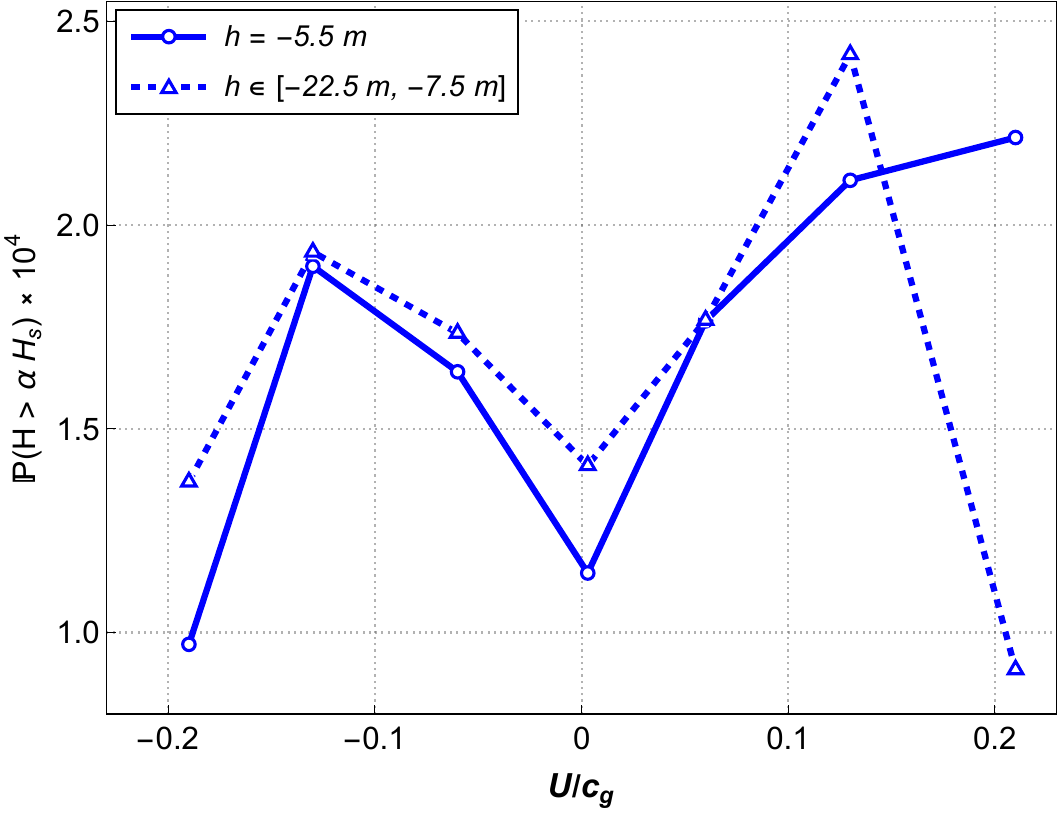}
\caption{Exceedance probability $P_\alpha$ for $\alpha = 2.0$ as a function of the normalized current velocity $U/c_g$ measured at $h=-5.5$~m and taken as an average in the range $-22.5\textrm{~m} < h < -7.5\textrm{~m}$.}
\label{fig:Udepth}
\end{figure}
We further investigate the behavior in different ranges of wave steepness. \jfm{Figure}~\ref{fig:Ualpha3} displays the rogue wave amplification probability as a function of the current speed, for three ranges of steepness, corresponding to linear ($\varepsilon \le 0.05$), second-order ($0.05 \le \varepsilon \le 0.09$) and higher-order ($\varepsilon \ge 0.9$) regimes, respectively. The amplification of the rogue wave probability observed in \jfm{figure}~\ref{fig:URogueFisher} mainly originates from the second-order steepness range. In contrast, the amplification is weaker in the linear regime, and vanishes in the higher-order regime. In the latter case, the onset of wave breaking associated with the large steepness likely levels off the wave height distribution, limiting the occurrence of rogue waves. 
Thus, both the homogeneity of a dataset and the steepness regime are relevant for rogue wave occurrence.

\subsection{Robustness assessment: Depth and rest conditions definitions}\label{section:SensitivityDepth}

\begin{figure}
\centering
    \includegraphics[scale=0.47]{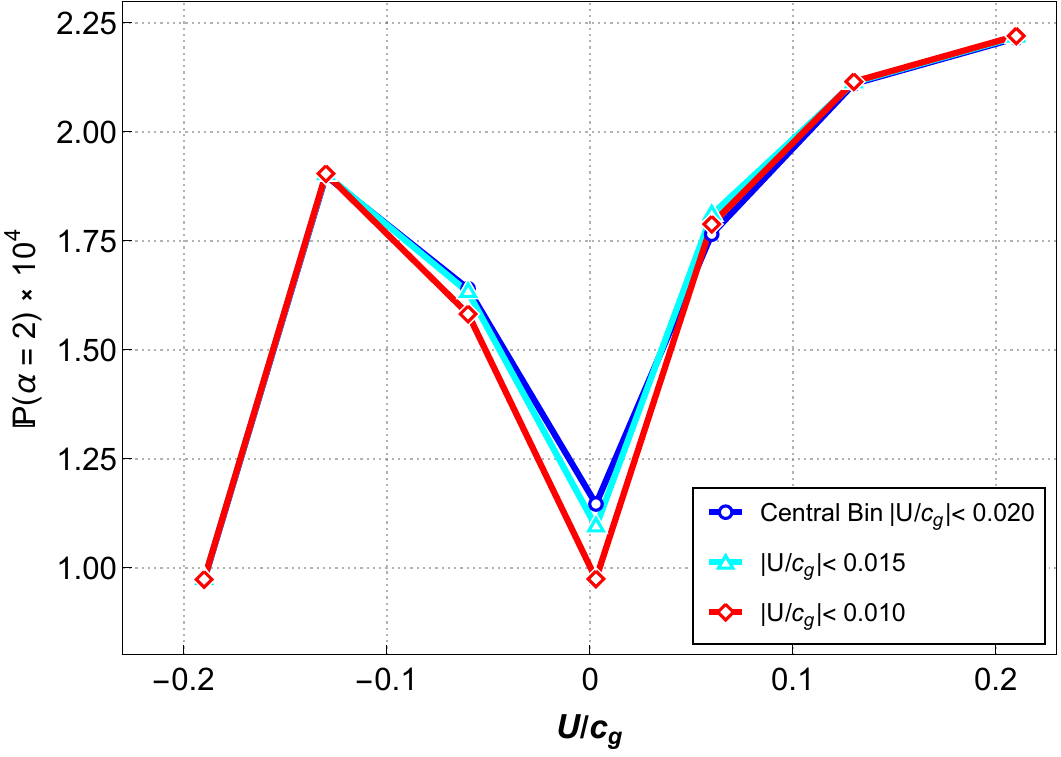}
\caption{Sensitivity to different definitions of the rest condition (legend) of the exceedance probability of rogue waves ($\alpha = 2$). }
\label{fig:U0DefRogue}
\end{figure}
As mentioned in \jfm{section} \ref{sec:methods}, the water depth at the measurement location (30~m) is sufficiently small compared to the wavelength ($\sim$~100~m) for the waves to be influenced by the current over the full water column. We checked that our results are robust against the choice of the depth at which the current speed was measured, by comparing our results obtained by considering the current at $h=-5.5$~m with those obtained by considering $U/c_g$ averaged between $h = -7.5$~m and $-22.5$~m. The behaviors were comparable, except in the rightmost bin corresponding to a fast following current (\jfm{figure}~\ref{fig:Udepth}). However, as discussed above, the extremely large confidence interval on this bin limits the relevance of these differences. A similar check was performed by considering the current at $h = -15$~m (not shown), leading to the same conclusion.

This relative independence of our results of the depth at which the current speed was measured, facilitates the comparison with experiments, where the vertical current profile is generally homogeneous~\citep{Ducrozet2021,Benoit2023}. Finally, \jfm{Figure}~\ref{fig:U0DefRogue} displays the rogue wave amplification probability for three definitions of the boundaries of the central bin, i.e., the rest condition. Obviously, the impact of this choice on our results is minimal, especially when comparing the differences between the curves with the width of the confidence interval of the data (see figure~\ref{fig:URogueFisher}).

\section{Discussion}

The most striking result from our observations is that both following and opposing currents amplify the rogue wave probability, compared to rest conditions. This surprising symmetrical behavior cannot be explained by the recently reported symmetric rise of the wave height with stronger currents \citep{TeutschSMJK2023}, because the rogue wave probability does not depend on $H_s$~\citep{Stansell2004,Christou2014,Mendes2021a}.

The strong reported increase in the relative water depth $k_ph$ for $|U/c_g| > 0.1$ as compared to rest conditions \citep{TeutschSMJK2023} cannot affect the wave statistics either, because the wave field is already in the deep water regime (\jfm{figure}~\ref{fig:spectraldist2}\jfm{b}). This absence of effect is well described by a spectral analysis of out-of-equilibrium systems \citep{Mendes2021b,Mendes2022,Mendes2023}. 
 
The steepness also increases, as compared to rest conditions, for both directions of current, but significantly more for an opposing one (see \jfm{figure}~\ref{fig:spectraldist2}\jfm{a} and figures 5a and 6a of \cite{TeutschSMJK2023}). This increase is expected to translate into an increase of the exceedance probability~\citep{Tayfun1980,Mendes2021b,Mendes2022,Mendes2023}. However, the asymmetry of the steepness growth prevents it from being the dominant driver of the symmetric increase of the exceedance probability.

As compared to the rest condition, currents increase the directional spread (Figure~6f in \cite{TeutschSMJK2023}) and decrease the bandwidth (Figure~6e in \cite{TeutschSMJK2023}). These dependencies, which are more marked for ($U/c_g \ge 0.08$), i.e. beyond the peaks of panels c and d of \jfm{figure}~\ref{fig:spectraldist2}, should respectively decrease~\citep{Mori2011, Mori2023} and increase~\citep{Higgins1980} the rogue wave probability, so that their effects likely compensate each other. 

Finally, large directional spread and bandwidth limit the Benjamin-Feir index (BFI $\propto \varepsilon / \nu$) to values insufficient to induce the large observed exceedance probability amplification. Furthermore, these two variables slowly change with increasing current speed, while the growth of the steepness is asymmetrical. Thus, the BFI can explain neither the magnitude of the amplification nor the symmetry between forward and opposing currents. 
 
The remaining variable capable of leading the rogue wave amplification is therefore the relative current speed $U/c_{g}$ itself. It was already pointed out theoretically by \citet{Toffoli2010,Toffoli2015} from the perspective of the NLSE framework that maximum amplitudes of regular waves should depend on $U/c_{g}$.
\citet{Mendes2022} have theoretically shown that the set-down of waves traveling past a shoal mostly controls the magnitude of the anomalous statistics. Laboratory experiments have demonstrated 
small changes in the wave-driven set-down due to wave-current interaction \citep{Svendsen1986,Jonsson2005}, but stronger ones for the effect of tides over reefs \citep{Becker2014,Yao2020,Yao2023}. The same should also apply in the open sea over flat bottoms. Indeed, the wave-driven set-down is negligible in deep water \citep{Higgins1962}, while the set-down due to wave-current interaction in deep water is proportional to the ratio $-U^{2}/2g$, which can be rewritten in terms of $-(U/c_{g})^{2}/8k$ \citep{Brevik1978,Jonsson1978}. This quadratic dependence on current velocity, hence on the associated set-down, may explain the observed symmetry of extreme wave statistics amplification.

Our finding regarding the effect of opposing and following currents of the same speed on rogue wave probability recalls the results of \citet{Waseda2015}, who have shown that the effect of currents on resonant interactions is symmetrical, although the main driving parameter was the current gradient. Investigating this aspect would however require long-term series from several locations, which is clearly out of scope of the present data and the analysis thereof.

The second important finding of the present work is the saturation of the extreme wave amplification for strong currents. Opposing currents will induce partial wave blocking and wave breaking~\citep{Wu2004,Toffoli2010b}, both likely boosting the wave steepness $k_{p}H_{s}$ up to a maximum of $0.18$ or $0.2$~\citep{Toffoli2010}, which we likely observe as the maximum rogue wave amplification around $|U/c_{g}| \approx -0.13$. In the case of following currents, however, $U > 0$, so that the sum $U+c_{g} > 0$: wave blocking cannot occur \citep{Unna1942}. It is therefore surprising that the amplification of rogue wave probability also saturates for following currents.

Furthermore, the tidal current oscillates with  a period of approximately 12~hours. Investigating the influence of this slight non-stationarity, as well as of spatial inhomogeneities on the extreme wave statistics \citep{Pizzo2023,Benetazzo2024} would provide further refinement of the comparison with flume experiments \citep{Toffoli2010,Toffoli2013,Waseda2015,Ducrozet2021}. 

\section{Conclusion}

In this work, we present the first long-term observational statistical study of the effect of both following and opposing currents on rogue wave statistics, based on data of wave-tide interaction. Following currents amplify rogue wave occurrence as much as opposing currents. The amplification reaches a maximum for $U/c_g \approx \pm 0.13$, and saturates for stronger currents of either direction. Therefore, not only waves traveling against large ocean currents such as the Gulf or Agulhas Stream are prone to rogue wave formation, but also waves interacting with tidal streams. This is a significant finding, shedding light on another type of risk. While strong wide streams are not that common, tidal currents are found on nearly every coast.

Moreover, subsets of the data set with different ranges of steepness display the same symmetry, but feature different magnitudes: second-order seas imply the highest amplification, followed by steep seas near breaking conditions, while linear seas have the lowest amplification magnitude.

Our results are comparable with experimental observations in unidimensional irregular wave fields interacting with stationary opposing currents. This similarity is striking, since our data are broad-banded and show a substantial directional spread, which are both known to weaken rogue wave occurrence \citep{Higgins1980,Karmpadakis2022}. However, when we narrow the data to smaller intervals of steepness, directional spread or a combination thereof, we see a twofold higher amplification in extreme wave statistics. Furthermore, we also observe that the steepness, and even more the directional spread, have the largest impact on the amplification by a current. This larger amplification in the homogeneous sample as compared to the full one is due to a combination of a higher exceedance probability in the presence of a current and a lower probability in rest conditions.

\section{Acknowledgements}
The authors are grateful for fruitful discussions about the physics of wave-current interactions with Prof. Takuji Waseda from the University of Tokyo and Prof. Alessandro Toffoli from the University of Melbourne. S.M and J.K. were supported by the Swiss National Science Foundation under grant 200020-175697.

The measurement data were collected and made freely available by the BSH marine environmental monitoring network (MARNET), the RAVE project (www.rave-offshore.de), the FINO project (www.fino-offshore.de) and cooperation partners of the BSH. The sea state portal was realized by the RAVE project (Research at alpha ventus), which was funded by the Federal Ministry for Economic Affairs and Climate Action on the basis of a resolution of the German Bundestag.

\textbf{Declaration of Interests}. The authors report no conflict of interest.

\bibliography{Maintext}

\end{document}